\documentclass[aps,prc,twocolumn,superscriptaddress,
amsfonts,amsmath,
amssymb,showpacs,floatfix]{revtex4}
\usepackage{graphicx,dcolumn}

\usepackage{graphicx,epsfig}
\usepackage{dcolumn} 
\usepackage{amsmath}

\begin{document}

\title{Nuclear Structure Relevant to Neutrinoless Double $\beta$ Decay: 
  $^{76}$Ge and $^{76}$Se.}

\author{ J.P.~Schiffer}
\email[correspondence to:]{schiffer@anl.gov}
\affiliation{Physics Division, Argonne National Laboratory, Argonne, IL 60439}
\author{S.J.~Freeman} 
\affiliation{University of Manchester, Manchester M13 9PL,U.K. }
\author{J.A.~Clark}
\affiliation{Yale University, New Haven, CT 06520}
\author{C.~Deibel}
\affiliation{Yale University, New Haven, CT 06520}
\author{C.R.~Fitzpatrick}
\affiliation{University of Manchester, Manchester M13 9PL,U.K. }
\author{S.~Gros}
\affiliation{Physics Division, Argonne National Laboratory, Argonne, IL 60439}
\author{A.~Heinz}
\affiliation{Yale University, New Haven, CT 06520}
\author{D.~Hirata}
\affiliation{GANIL (IN2P3/CNRS -DSM/CEA), B.P. 55027 14076 Caen Cedex 5, France}
\affiliation{The Open University, Dept. of Physics and Astronomy, Milton 
  Keynes, MK7 6AA, U.K.}
\author{C.L.~Jiang}
\affiliation{Physics Division, Argonne National Laboratory, Argonne, IL 60439}
\author{B.P.~Kay}
\affiliation{University of Manchester, Manchester M13 9PL,U.K. }
\author{A.~Parikh}
\affiliation{Yale University, New Haven, CT 06520}
\author{P.D.~Parker}
\affiliation{Yale University, New Haven, CT 06520}
\author{K.E.~Rehm}
\affiliation{Physics Division, Argonne National Laboratory, Argonne, IL 60439}
\author{A.C.C.~Villari}
\affiliation{GANIL (IN2P3/CNRS -DSM/CEA), B.P. 55027 14076 Caen Cedex 5, France}
\author{V.~Werner}
\affiliation{Yale University, New Haven, CT 06520}
\author{C.~Wrede}
\affiliation{Yale University, New Haven, CT 06520}

\date{\today} \begin{abstract} 
 The possibility of observing neutrinoless double $\beta$ decay offers the
 opportunity of determining the neutrino mass $\it if$ the nuclear matrix
 element were known.  Theoretical calculations are uncertain and
 measurements of the occupations of valence orbits by nucleons active in the
 decay can be important.  The occupation of valence neutron orbits in the
 ground states of $^{76}$Ge and $^{76}$Se were determined by precisely
 measuring cross sections for both neutron-adding and removing transfer
 reactions. Our results indicate that the Fermi surface is much more diffuse
 than in theoretical (QRPA) calculations. We find that the populations of at
 least three orbits change significantly between these two ground states
 while in the calculations the changes are confined primarily to one orbit.
\end{abstract}

%
\maketitle 

An essential step in studying the nature of the neutrino is the attempt to
observe neutrinoless double beta decay \cite{Vogel} and major efforts are
being undertaken with this objective in mind.  Observation of such a process
would immediately show that neutrinos are their own antiparticles, and its
rate may well give the first direct measure of the neutrino mass {\it if}
the corresponding nuclear matrix element can be reliably calculated. As an
example, for one of the likely candidates ($^{76}$Ge), theoretical
calculations have yielded answers that are spread over more than an order of
magnitude.  This prompted the statement by Bahcall $\it{et~al.}$
\cite{Bahcall} {\it ``The uncertainty in the calculated nuclear matrix
elements for neutrinoless double beta decay will constitute the principal
obstacle to answering some basic questions about neutrinos''}.  There have
been suggestions that relate the neutrinoless double-beta-decay matrix
elements to those for ordinary single beta decay, or to the `normal'
two-neutrino modes which have been observed experimentally \cite{Suhonen}.
However, neutrinoless decay proceeds by the virtual excitation of states in
the intermediate nucleus with a momentum transfer much larger than that for
these other processes. It will thus involve all possible virtual
intermediate states (up to about 100 MeV of excitation), and so will include
giant resonances.  There is no other experimentally accessible process
that could directly determine the matrix element.

Although there is still considerable discussion regarding the best
theoretical approach, what unquestionably matters is knowing the population
of the valence orbits for the nucleons that switch from neutrons to protons.
We have therefore undertaken a set of measurements to determine this
quantity experimentally, and report here on the valence neutron populations
and the differences in these populations for $^{76}$Ge and $^{76}$Se.  In a
previous experiment we determined that the neutron pair correlations in
these two nuclei are quantitatively very similar \cite{gept}. 

The Macfarlane-French \cite{macfarlane} sum rules for nucleon transfer state
that the summed spectroscopic strength for neutron-{\em adding} reactions
with a given set of quantum numbers is equal to the vacancies in that target
orbital, while the sum over states for neutron-{\em removing} reactions will
determine the occupancy.  Here we have measured the cross sections and
extracted spectroscopic factors of significantly populated states, for both
neutron-adding and neutron-removing reactions.  The summed spectroscopic
factors for $\it{both}$ reactions can be added and used to provide a
normalization, allowing occupation numbers for orbitals to be extracted.

The nucleon transfers reported here have been measured previously
\cite{germanium,selenium}, but not with the same experimental methods, and
using different parameters in each DWBA analysis for extracting
spectroscopic factors. The aim of the present measurement is to analyze all
the results in a consistent manner to permit the extraction of more accurate
occupation numbers with a common experimental approach.

The active orbits for neutrons in these nuclei, with 42 and 44 neutrons, are
$1p_{3/2}$, $0f_{5/2}$, $1p_{1/2}$, and $0g_{9/2}$.  We have made systematic
measurements to obtain accurate cross sections for the neutron-adding (d,p)
and ($\alpha,^3$He) reactions, as well as for neutron-removing (p,d) and
($^3$He,$\alpha$) reactions.  The momentum matching in (d,p) reactions for
transitions with $\ell$=3 and 4 is not optimal and thus the cross sections
are rather weak. Therefore helium-induced reactions were used to obtain data
with improved momentum matching and larger cross sections for the
higher-$\ell$ transitions. This selectivity is illustrated in Figure 1.

Deuteron, proton, alpha, and $^3$He beams from the Yale tandem accelerator
were used to bombard isotopically enriched Ge and Se targets of about
200-300 $\mu$g/cm$^2$ evaporated on thin, 50 $\mu$g/cm$^2$ C foils. The
momenta of the reaction products were determined and the particles
identified with the Yale Enge spectrograph and gas-filled focal-plane
detector backed by a scintillator.

The product of target thickness and spectrometer solid angle was found by
measuring elastic scattering in the Coulomb regime at 30$^o$ for each target
used.  The beam energies used for this calibration were 6-MeV protons and
10-MeV alphas.  For the transfer reactions, the same spectrometer aperture and
beam integrator settings were used to minimize potential systematic errors.
The beam energies chosen were 15 MeV for the (d,p) reaction and 23 MeV for
the (p,d) to keep the energies in each channel comparable. Similarly,
($\alpha,^3$He) was studied at 40 MeV and ($^3$He,$\alpha$) at 26 MeV.
Measurements were also carried out on targets of $^{74}$Ge and $^{78}$Se to
provide an additional check. The energy resolution obtained was $\sim$40 keV
for the deuteron and proton-induced reactions, and $\sim$70 keV for the
$^{3,4}$He reactions.

\begin{center}    
\begin{figure}
\includegraphics[scale=0.35,angle=270]{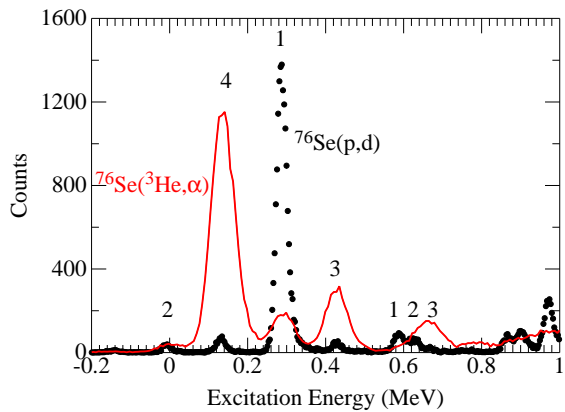}%
\caption
{Energy spectra for the neutron-removal reactions for $^{76}$Se to $^{75}$Se
illustrating how different $\ell$ values are favored by different momentum
transfers in the two reactions. The $\ell$=1 transitions appear strongly in
the 11$^o$ (p,d) spectrum (points) while the $\ell$=3 and, in particular,
$\ell$=4 are most prominent in ($^3$He,$\alpha$) (line) where the resolution
is worse because of the higher energy. The $\ell$ values are indicated in
numbers above the peaks.  }
\label{Figure1.}
\end{figure}
\end{center}
The (d,p) angular distributions have been studied previously and $\ell$
values were assigned \cite{germanium,selenium}.  In the current work, the
yields were therefore measured only at the angles that correspond to the
peaks in the angular distributions for the $\ell$ values of interest:
11$^o$, 28$^o$ and 37$^o$ for $\ell$=1, 3 and 4 respectively. The
helium-induced reactions are forward peaked and so the most practical
forwardmost angles were chosen: 5$^o$ for ($\alpha,^3$He) and 8$^o$ for its
inverse.  The previous $\ell$-value assignments \cite{germanium,selenium}
were confirmed, as may be seen in Figure 2.  Our results also agree
approximately with the previous relative spectroscopic factors for states
populated with a particular target.
\begin{center}
\begin{figure}
\includegraphics[scale=0.5,angle=0]{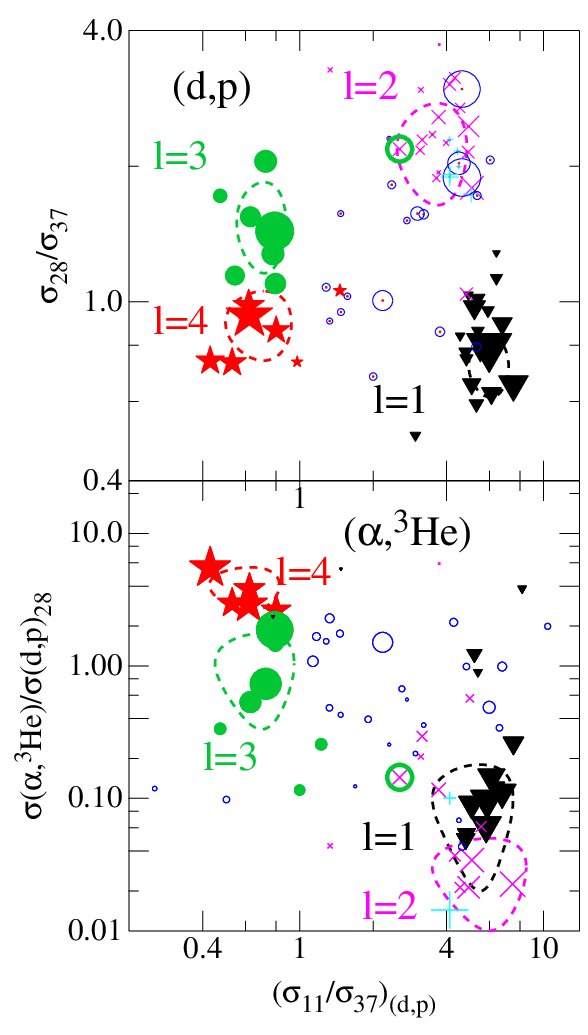}%
\caption {Ratios of cross sections, $\sigma_{d,p}(28^o)/\sigma_{d,p}(37^o)$
$\it{vs.}$ $\sigma_{d,p}(11^o)/\sigma_{d,p}(37^o)$ on top and
$\sigma_{\alpha,^3He}/\sigma_{d,p}(28^o)$ $\it{vs.}$
$\sigma_{d,p}(11^o)/\sigma_{d,p}(37^o)$ below, are shown for different
$\ell$ values and reactions. The symbols, one for each state, indicate the
$\ell$-value assignments from previous work: triangles (black) are $\ell$=1,
circles (green online) are $\ell$=3, and stars (red online) are $\ell$=4. In
addition, states not included in the analysis are $\ell$=2 transitions
indicated by $\times$ and $\ell$=0 by + signs. States with unknown $\ell$
values are indicated by hollow circles (in blue online). The size of the
symbols is a rough measure of the cross sections. The dashed lines indicate
the loci of the ratios for well-established $\ell$ values. The $\times$
surrounded by a (green online) circle, between the $\ell$=2 and 3 islands in
the lower box, is the 500-keV $5/2^+-5/2^-$ doublet in $^{77}$Ge discussed
in the text.}
\label{Figure 2.}
\end{figure}
\end{center}

We used the finite-range code PTOLEMY \cite{ptolemy} for the DWBA
calculations. The normalization depends somewhat on the choice of distorting
parameters.  The extracted relative spectroscopic factors also vary slightly
and this is a source of some of the uncertainty at the level of a few \%.
For the projectile bound-state wave function, the Reid potential was used
for the deuteron, and a Woods-Saxon one for the $\alpha$ particle and for
the various target bound states. 

Absolute spectroscopic factors are notoriously difficult to obtain. The
values of spectroscopic factors for `good' single-particle states in
doubly-magic nuclei are usually around 0.5 - 0.6 because of short-range
correlations. Such correlations are expected to be a uniform property of
nuclei, not changing between nearby nuclei or configurations. Since the
overall effect in depleting absolute strength is expected to be uniform, it
can be compensated for by a renormalization of strength and examining the
relative strengths of spectroscopic factors through the sum rules. Since the
sums of the strengths for neutron adding or removing are proportional to the
vacancies or occupancies, together they should add up to the $(2J+1)$
degeneracy of the orbits and can serve as such a normalization. A check is
provided, in that the summed spectroscopic factors for a given orbit should
add up to the {\it same} value for each of the targets.  The mean
normalization factors with the potentials adopted for the $\ell$=1, 3, and 4
transitions are 0.53, 0.56, and 0.57 respectively with rms fluctuations
among the targets of 2, 12, and 7 \%, indicating that the procedure is
reasonable. This normalization constant for the two reactions is somewhat
surprisingly close to the depletion that should be expected for `absolute'
spectroscopic factors.

Several points are to be noted in the above sums.  Since not all the spins
of the states seen in $\ell$=1 transitions are known, we summed all $\ell$=1
transitions, thus combining the $j$=1/2 and 3/2 states. For the
neutron-removal reactions a small correction was made for the unobserved
$T_{>}$ isobaric analog states, corresponding to proton removal. Also for
neutron removal, in the (p,d) reaction, some previously determined $\ell$=1
transitions at high excitation energy were beyond the energy range measured
here. A correction was made for these states by normalizing the previously
determined spectroscopic factors to ones determined in the present work.
There were no known missed states for the neutron-adding measurements or for
the other $\ell$ values.

Finally, for the $f_{5/2}$ states, no $5/2^{-}$ state was known in
$^{77}$Ge, while all other nuclei in this region have such a state well
below 1 MeV in excitation energy.  In attempting to find such a state in the
($\alpha,^3$He) reaction, the intensity of the peak around 500 keV
excitation was stronger than expected for a known $\ell$=2 transition to a
$5/2^{+}$ state at 504.8 keV, but the centroid of this peak seemed lower
than the accepted value -- around 492 keV. In fact, a tentative state is
reported in the compilations \cite{ENSDF} at 491.9 keV from unpublished work
with the ($^{13}$C,$^{12}$C) reaction, and we have assumed that this is the
missing $5/2^{-}$ state. Its strength was included in the sums.

The vacancies and occupancies from the summed normalized spectroscopic
factors are shown in Table I.  Listed in the Table are the numbers of holes
and particles from neutron adding and removing, their sum, and the best
average value of the occupancy, all computed with a constant normalization.
The $\ell$=1 strength is best determined in the (d,p) and (p,d) reactions
and the $\ell$=3 and 4 transitions from the helium-induced reactions. As was
noted, the sums of holes and particles for both $\ell$=1 and 4 transfers are
constant to better than 5\% across the targets studied. For $\ell$=3, the
situation is somewhat worse, partly because these transitions are relatively
weak in both reactions.  As a result, components of strength could have been
missed. Additionally, there is some ambiguity about the $\ell$=3 strength in
the pickup reactions since some of the transitions could be to $7/2^{-}$
states at higher excitation energies. For cases where there is no evidence
on the spins of higher $\ell$=3 hole states we, somewhat arbitrarily,
excluded all $\ell$=3 transitions above 1 MeV excitation.  As was noted, the
summed strengths for $\ell$=3 fluctuate more than the others.

Our measurements provide two determinations of the valence-orbit occupancies
in the $^{76}$Ge and $^{76}$Se ground states, one from the neutron-adding
data, and one from the neutron removal. We average these, weighting the
former by a factor of two.  There are two reasons for this.  First, the
major neutron shell between N=28 and 50 is more than half filled, with about
twice as many particles as holes. A given fractional error therefore leads
to a bigger uncertainty in the particle number compared to the number of
holes, at least for $\ell$=1 and 3. Second, the sensitivity of the
calculated DWBA cross sections to distorting parameters is larger for the
($^3$He,$\alpha$) reaction ($\approx$15-25\%) than for its inverse
($\approx$0-10\%).

 \begin{table} 
    \caption{Summed spectroscopic strengths.}
	 
\label{one} \begin{tabular}{|l|c|c|c|c|r|} \hline
  Target& {\rm Holes}&  {\rm Particles} &{\rm Holes +} &{\rm Adopted}\\
  & & & {\rm Particles} & {\rm Occupancy} \\\colrule
$^{74}$Ge $\ell$=1 & 1.15 &        &       &  4.85 \\
$^{76}$Ge          & 1.12 & 4.83   &  5.95 & $ \bf{4.87\pm0.20}$  \\
$^{76}$Se          & 1.63 & 4.49   &  6.12 & $ \bf{4.41\pm0.20}$  \\
$^{78}$Se          & 0.94 &        &       & 5.06  \\
\hline
$^{74}$Ge $\ell$=3 & 1.90 & 4.38 &  6.28 & 4.19  \\
$^{76}$Ge          & 1.14 & 3.92 &  5.06 & $\bf{4.56\pm0.40}$  \\
$^{76}$Se          & 2.10 & 3.71 &  5.81 & $\bf{3.83\pm0.40}$  \\
$^{78}$Se          & 2.34 & 4.63 &  6.97 & 3.98    \\
\hline
$^{74}$Ge $\ell$=4  & 4.37 & 5.83 & 10.20 & 5.69  \\
$^{76}$Ge           & 3.41 & 6.27 &  9.68 & $\bf{6.48\pm0.30}$     \\
$^{76}$Se           & 4.36 & 6.13 & 10.49 & $\bf{5.80\pm0.30}$     \\
$^{78}$Se           & 2.80 & 7.31 & 10.11 & 7.24     \\
\hline
\colrule \end{tabular} \end{table}

The uncertainties in the final mean occupancy values are difficult to
estimate. Statistical errors in the summed strength are less than 1\% and
relative systematic errors between targets are believed to be less than 3\%.
The biggest uncertainties stem from possible missed states, especially for
the $\ell$=3 transitions, and from uncertainties in the DWBA calculations.
We estimate that the occupancy is determined to about 0.2 nucleons for the
$1p$, 0.3 for the $0g_{9/2}$ orbits, and slightly worse, 0.4, for the
$0f_{5/2}$ orbit. These estimates of uncertainties are rather crude.
However, there are several checks that give us some confidence:

\begin{itemize}
 \item The normalization factors obtained for each target separately are
       similar.
 \item The mean normalizations for each $\ell$ value and reaction type are
       also similar.
 \item The summed removing and adding strengths for $^{74,76}$Ge
       and $^{76,78}$Se, 22.5, 20.7, 22.4, and 23.1 respectively, are
       consistent with the expected value of 22.0.
 \item As an independent result, the neutron vacancies obtained for the
       four nuclei (from the adopted occupancies in the Table) are 7.3, 6.1,
	  7.9 and 5.7, in good agreement with the expected values of 8, 6, 8,
	  and 6.
\end{itemize}

Beyond the valence $1p$, $0f_{5/2}$ and $0g_{9/2}$ orbitals, neutron removal
from $^{76}$Se suggests that approximately 0.2 neutrons are in the
$1d_{5/2}$ orbit. The weak $5/2^+$ state in $^{75}$Ge is not resolved in our
work, but using the results in \cite{ENSDF}, we obtain a roughly similar
value.  
\begin{center}
\begin{figure}
\includegraphics[scale=0.5,angle=0]{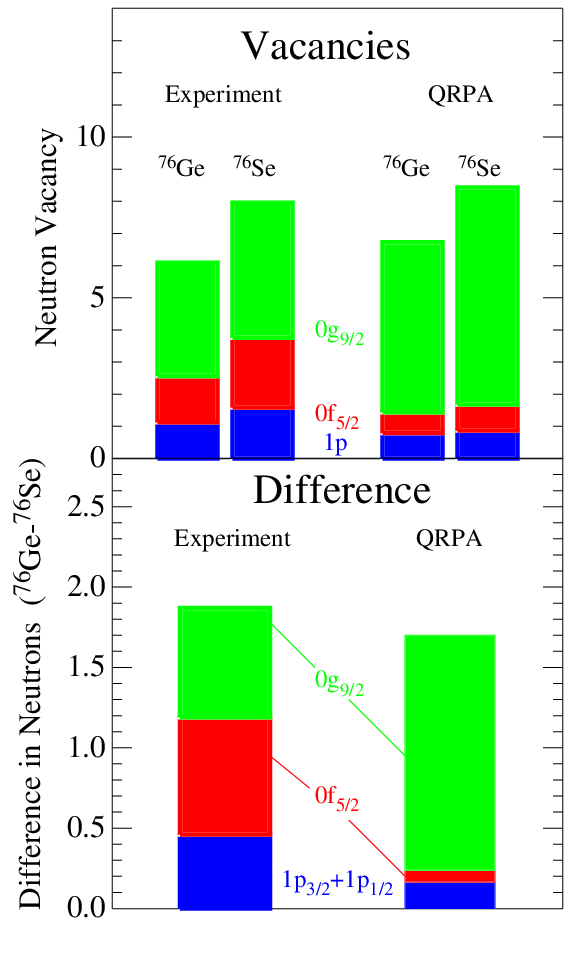}%
\caption
{The deduced neutron vacancies for $^{76}$Ge and $^{76}$Se are shown in the
three active valence orbits and compared to those from the QRPA calculations
of Reference 10.  The naive shell closure should give 6 and 8 vacancies for
these two nuclei. The lower part of the figure shows the $\it{differences}$
in these occupations (expected to be 2.0), again compared to the QRPA
calculation.}
\label{Figure 3.}
\end{figure}
\end{center}

The values of vacancies are shown in Figure 3 along with the QRPA results
\cite{Rodin}.  There is little question that the vacancies in the $1p$ and,
especially, in the $0f_{5/2}$ orbits are significantly larger in the data
than in the calculations. For the neutrinoless double-beta-decay experiments
it is the ${\rm changes}$ in occupancy that are important, and so in the
lower part of Figure 3 we show the differences between $^{76}$Ge and
$^{76}$Se: 0.46$\pm0.20$ in $1p$, 0.73$\pm$0.40 in $0f_{5/2}$ and
0.68$\pm$0.30 in $0g_{9/2}$.

While the QRPA results predict changes between the two nuclei to be mostly
in the $0g_{9/2}$ orbit, the experiment shows quite clearly that the changes
in the $1p$ and $0f_{5/2}$ orbits are much larger than predicted. The
qualitative feature that, in disagreement with QRPA, there are still large
vacancies in $1p$ and $0f_{5/2}$ is quite robust. It follows from the
relatively large cross sections in the neutron-adding reactions and it
cannot depend on the details of the analysis or the assumptions.

What the consequences may be, of this disagreement in neutron occupancy
between QRPA and experiment, on the matrix element for neutrinoless double
beta decay are not clear at present and will need to be investigated in more
detail. Proton occupancies are similarly important and experiments to
determine them are planned.

We are indebted to V.A. Rodin and A. Faessler for sending us the results of
the QRPA calculations.  We wish to acknowledge the help of the operating
staff of the Yale tandem and of John Greene at Argonne for meticulous work
on target preparation. The work was supported by the U.S. Department of
Energy, Office of Nuclear Physics, under contracts DE-FG02-91ER-40609 and
DE-AC02-06CH11357, the UK Science and Technology Facilities Council, and
IN2P3/CNRS-France.

\end{document}